\documentclass[aps,prl,twocolumn,showpacs]{revtex4}
\usepackage{amsfonts,amssymb,amsmath,latexsym,epsfig,color}
\def\Ptree{P_{\textrm{I}}}
\def\ztree{z_{\textrm{I}}}

\begin{document}

\title{On universality of algebraic decays in Hamiltonian systems}

\author{G. Cristadoro$^{1,2}$}
\author{R. Ketzmerick$^{3,4}$ }

\affiliation{$^{1}$ Max Planck Institute for the Physics of Complex Systems, N\"othnitzer Str. 38, 01187 Dresden, Germany\\
$^{2}$  Center for Nonlinear and Complex Systems, Via Valleggio 11, 22100 Como, Italy \\
$^{3}$ Institut f\"{u}r Theoretische Physik, Technische Universit\"at Dresden, 01062 Dresden, Germany\\
$^4$ Kavli Institute for Theoretical Physics, UCSB, Santa Barbara, CA93106}

\date{\today}

\begin{abstract}

Hamiltonian systems with a mixed phase space typically exhibit an algebraic decay of correlations and of Poincar\'e recurrences, with numerical experiments over finite times showing system-dependent power-law exponents. We conjecture the existence of a universal asymptotic decay based on results for a Markov tree model with {\it random} scaling factors for the transition probabilities.
Numerical simulations for different Hamiltonian systems support this conjecture and permit the determination of the universal exponent.
 
\end{abstract}


\maketitle

The phase space of  Hamiltonian systems with two degrees of freedom generally shows an intricate mixture of chaotic and regular structures.  Regular regions consisting of  `islands' of  quasi-periodic motion (KAM tori) appear hierarchically  interspersed in a chaotic sea. The character of the motion in the irregular component is dominated by stickiness of trajectories close to the boundary  circles separating regular from irregular regions.
Stickiness crucially affects statistical quantities, like correlations, the distribution of Poincar\'e recurrences  or anomalous diffusion, which show related algebraic behaviors.
Finite-time numerical experiments revealed system-dependent values for the power-law exponent $z$ of the decay of the distribution of Poincar\'e recurrences, $P(t)\sim t^{-z}$, that is the probability to return to a given region 
after a time larger than t.
Great effort was devoted over the last two decades to understand if there is an asymptotic universal decay for $P(t)$ and how its temporal behavior 
is related to the hierarchical phase-space structure near regular boundaries 
\cite{ChiShe81, Kar83, MacMeiPer84, ChiShe84, HanCarMei85, MeiOtt85+86, GraKan85, ZasEdeNiy97, ChiShe99, WeiHufKet03}.
In particular, renormalization techniques were used to derive scaling relations  for the self-similar structure close to the critical circle with golden mean frequency \cite{HanCarMei85, ChiShe99} or for trapping in exact self-similar island-around-island structures \cite{ZasEdeNiy97}. More recently it was demonstrated~\cite{WeiHufKet03} that both scenarios are usually present in an Hamiltonian system, thus strengthening the validity of a Markov tree approach~\cite{MeiOtt85+86}.
While the self-similar Markov tree model succeeds to reproduce a power-law decay, it is based on the approximation of exact self-similarity~\cite{MeiOtt85+86,GreMacSta86}. It remains an open question how deviations from exact scaling affect the algebraic decay and the issue of universality.

\begin{figure}[t!]
\centerline{\epsfxsize=\columnwidth  \epsfbox{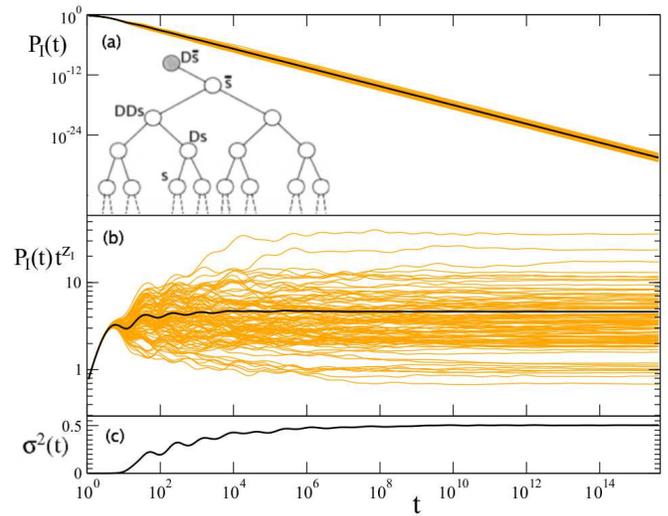}}
\vglue -0.4cm
\caption{(Color online) 
(a) Survival probability $\Ptree(t)$ for $100$ realizations of random scaling factors (thin lines) and their logarithmic average (thick line). 
Inset: binary tree with root $\bar{s}$ and absorbing site $D\bar{s}$.
(b) Rescaled survival probabilities  $\Ptree(t) \, t^{\ztree}$ with $\ztree$ from Eq.~(\ref{simplercondition}), showing the same asymptotic power-law exponents.
(c) Variance $\sigma^2(t)$ of $\ln [\Ptree(t) \, t^{\ztree}]$ showing saturation.
}
\label{fig1}
\end{figure}

In this paper we study the role of random scaling factors in a Markov tree model.
Initially, different realizations of the randomness give rise to quite different power-law behaviors  (see Fig.~\ref{fig1}). In contrast,  we show that asymptotically the exponent of the algebraic decay is the same for all realizations.
We argue that this is representative for Hamiltonian systems and present strong numerical support. As a consequence the average of $P(t)$ over different  Hamiltonian systems  is a meaningful quantity  and allows for an efficient determination of the asymptotic universal decay. We numerically extract a universal exponent $z \simeq 1.57$ for Poincar\'e recurrences in Hamiltonian systems. 
\newline

We consider a Markov model on a binary tree. For an arbitrary site  $s$, we call $Ds$  the  site obtained by going up one level in the tree towards the root site $\bar{s}$ (see Fig.~\ref{fig1}a, inset).  The dynamics on the tree is defined by the probabilities to move in one time step to one of the neighboring sites.  As we are interested in the survival probability we put $D\bar{s}$ as an absorbing site and we fix a non-zero  probability of leaving the tree $p_{\bar{s} \to D\bar{s}}$~\cite{footnote1}. The transition probabilities 
for all  sites $s$  below $\bar{s}$  are then chosen according to
\begin{eqnarray}\label{ourmodel}
p_{Ds  \to s }/p_{s \to Ds} &=&\, A_s \\
p_{s \to Ds}/p_{Ds  \to DDs} &=& \, B_s \nonumber
,
\end{eqnarray}
where the scaling factors $A_s$ and $B_s$ are positive random variables with the constraints (i) $\langle A \rangle <1/2$ and (ii) $B <1$. 
The constraint (ii) ensures that the transition probabilities decrease moving away from the root $\bar{s}$. The  constraint (i) allows to define an initial probability density $\rho_s(t=0)$, with normalization $\sum_{s} \rho_s(0)=1$, that would be invariant for the closed system ($p_{\bar{s} \to D\bar{s}}=0$) by satisfying detailed balance $\rho_s(0) p_{s \to Ds}=\rho_{Ds}(0) p_{Ds \to s }$. 
Due to the non-zero transition probability $p_{\bar{s} \to D\bar{s}}$ of leaving the tree towards the absorbing site $D\bar{s}$  there is a decaying survival probability 
\begin{equation}\label{sumdensities}
\Ptree(t):=\sum_{s} \rho_s(t)
,
\end{equation}
of being still inside the tree at time t, with the subscript~\textrm{I} standing for the  ``invariant'' initial condition. 

In order to quantify the decay of  the survival probability  we make use of the following observation: In the sum, Eq.~(\ref{sumdensities}), the density $\rho_s(t)$ at a site $s$ gives its most important contribution while it is decaying exponentially, with a rate that is dominated by the probability $p_{s\to Ds}$ to escape in upward direction. At later times, even though its decay becomes algebraically slow, its contribution can be neglected \cite{footnote2}.  The survival probability can thus be approximated by a sum of decaying densities \cite{BarGil02}
\begin{equation}\label{sum}
\Ptree(t) \approx \sum_s \rho_s(0) e^{-p_{_{s\to Ds}} t}
.
\end{equation}
Numerically, we find this to be a good approximation, in the sense that the ratio with the exact decay of Eq.~(\ref{sumdensities}) is found to be asymptotically constant. 

Power-law asymptotics can be conveniently extracted using singularity analysis of the Mellin transform. An algebraic decay $P(t) \sim t^{-\alpha}$ is transformed into a simple pole of its Mellin transform $P^*(z)=\int {\textrm{dt} \, t^{z-1}P(t)}$ at $z=\alpha$. In particular we find using Eq.~(\ref{sum})
\begin{eqnarray}\label{Mellin}
\Ptree^*(z)&=& \Gamma(z) \sum_s \frac{\rho_s(0)}{p_{s\to Ds}^{z}} \nonumber \\
&=&\Gamma(z)\frac{\rho_{\bar{s}}(0)}{p_{\bar{s}\to D\bar{s}}^z} \; \sum_s 
\; \prod_{i=0}^{n(s)-1}  \frac{A_{D^i s}}{B^z_{D^i s}} 
\end{eqnarray}
where $\Gamma(z)$ is the Gamma function and where we used
\begin{eqnarray}
\rho_s(0)&=& \rho_{\bar{s}}(0)\prod_{i=0}^{n(s)-1}A_{D^i s}\\
p_{s\to Ds}&=&p_{\bar{s}\to D\bar{s}}\prod_{i=0}^{n(s)-1}B_{D^i s}
.
\end{eqnarray}
The products run over the sites $D^is$ encountered on the direct path from $s$ to the root $\bar{s}$ and $n(s)$ is the number of levels ascended.

As we are interested in the asymptotic behavior of $\Ptree(t)$ for $t\to \infty$, we need to find the right boundary of the strip of convergence of Eq.(\ref{Mellin}). The Gamma function has no poles with positive real value and thus we concentrate on the factor containing the sum over the sites $s$.  Fixing $z$ real, the infinite sum of products of random variables $\frac{A}{B^z}$ diverges with probability one~\cite{KeaDwyBre97} if $\beta(z) > 1/2$, where
\begin{equation}\label{minimum}
\beta(z)=
\min_{0\leqslant\sigma\leqslant1} \left\langle \left(\frac{A}{B^{z}}\right)^ {\sigma} \right\rangle
.
\end{equation}
Using the constraints (i) and (ii) of our random variables $A$ and $B$, one finds that $\beta(z)$ is monotonically increasing from $\beta(z=0)<1/2$ to $\beta(z \to \infty)=1$. This implies that there is a critical $\ztree$ with
\begin{equation}\label{condition}
	\beta(\ztree)=1/2
.
\end{equation}
Thus the survival probability asymptotically decays as
\begin{equation}\label{averagepower}
 \Ptree(t) \sim C(t) t^{-\ztree}
,
\end{equation}
with $C(t)$ growing at most logarithmically \cite{footnote-mellin}.	
We stress that $\ztree$ depends on the probability distributions of the scaling factors $A$ and $B$ only  and not on the specific values appearing in a given tree. The power-law exponent $\ztree$ is thus realization independent.

\begin{figure}[t!]
\centerline{\epsfxsize=\columnwidth  \epsfbox{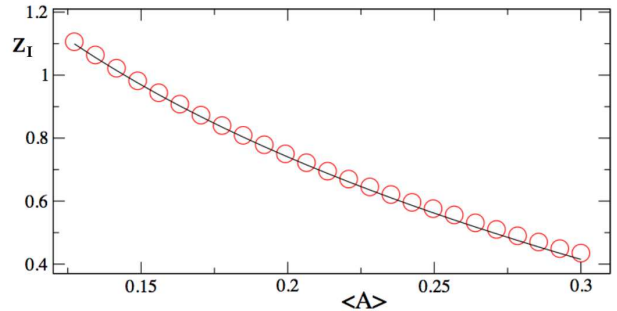}}
\vglue -0.5cm
\caption{(Color online)
Power-law  exponent $\ztree$ vs.\ $\langle A \rangle $ according to analytic estimate Eq.~(\ref{simplercondition}) (line) and from fitting the asymptotic decay of the numerically computed exact evolution, Eq.~(\ref{sumdensities}), for a single realization of scaling factors (circles). The distribution for the scaling factor $A_s$ ($B_s$) is chosen to be uniform on an interval centered around $\langle A \rangle$  ($\langle B \rangle=0.3$) with width $0.12$ ($0.1$) and is independent from $B_s$ ($A_s$).
}
\label{fig2}
\end{figure}

For not too wide distributions of $A$ and $B$, e.g.
$\langle (\frac{A}{B^{\ztree}})^2 \rangle < 2\;\langle \frac{A}{B^{\ztree}} \rangle^2$,
one can show that the minimum in Eq.(\ref{minimum}) appears at $\sigma=1$ and Eq.~(\ref{condition}) simplifies to
\begin{equation}\label{simplercondition}
\left\langle \frac{A}{B^{\ztree}} \right\rangle =1/2
.
\end{equation}
A numerical example is shown in Fig.~\ref{fig1} for random variables $A$ and $B$ independent and uniformly distributed on an interval in logarithmic scale with
$\ln A \in [-4.21,-3.17]$ and $\ln B \in [-2.02,-0.98]$, giving $\ztree=1.88$.
It clearly demonstrates that, even though initially quite different power laws emerge, asymptotically all realizations have the same power-law exponent. In addition, one observes that the averaged curve shows this asymptotic exponent beginning at much smaller times.

Moreover, in this case of not too wide distributions, one can show that fluctuations around the asymptotic algebraic decay vanish for each realization, as observed in Fig.~\ref{fig1}b. This  is due to an increasing number of sites $s$ contributing significantly to Eq.~(\ref{sum}) as time increases. We have numerically demonstrated this by finding that the inverse participation ratio $\sum_s [\rho_s(t)/P(t)]^2$ of the contributing densities  decreases like a power law as a function of time (not shown). It stresses the fact that the contribution to $P(t)$ at large times arises from an increasing number of paths~\cite{WeiHufKet03}.

In Fig.~\ref{fig2} we show numerical calculations of the power-law exponent fitted from the \emph{exact} evolution Eq.~(\ref{sumdensities}) for a \emph{single} realization of scaling factors, showing good agreement with the analytical prediction, Eq.~(\ref{simplercondition}). This agreement gives also an indirect confirmation of the validity of the approximation used in Eq.~(\ref{sum}).
We note, that Eq.~(\ref{simplercondition}) reproduces the power-law exponent of deterministically chosen scaling factors $A$ and $B$ in Ref.~\cite{MeiOtt85+86}.
\newline

We now want to argue why this model is able to capture the essential features of transport in Hamiltonian systems with a mixed phase space. The discrete nature of the Markov model is justified by the presence of invariant structures in  phase space that act as partial barriers (e.g.\ cantori with minimum flux) and that effectively separate
regions where the trajectory spends a long time before moving to the adjacent one. The specific set of transition probabilities depends  on the precise structure of the phase space, e.g., the transition probability for a trajectory to leave a region of area $a$ leaking through a partial barrier with flux $\Delta w$ is given by  $\Delta w/ a$. Area conservation assures that the flux is the same in both directions and thus the scaling factors $A_s$ in Eq.~(\ref{ourmodel}) are related to the scaling of areas of the connected regions, while the factors $B_s$ are related to the ratio of escape rates.

A typical mixed phase space is hierarchically organized~\cite{Mei92}. While approaching the boundary circle of a given island of regular motion (level-renormalization~\cite{Mac82}) there are secondary island chains encircling the main one (class-renormalization~\cite{Mei86}). This picture repeats on all scales ad infinitum.
Using these features of a mixed phase space Meiss and Ott~\cite{MeiOtt85+86} modeled the transport in Hamiltonian systems as a random walk on a binary tree with \emph{fixed} scaling factors for level- and class-scaling, respectively. However, typical scaling factors in class-scaling and, at least initially, in level-scaling do vary and a more realistic model must take into account these variations, as it was suggested in Ref.~\cite{MeiOtt85+86}.

Our proposed model with random scaling factors, Eq.~(\ref{ourmodel}), is the simplest of this kind. It is not aimed at giving a quantitative description, but it elucidates the qualitative consequences of randomness in the scaling factors, namely the presence of a well-defined asymptotic algebraic decay with a realization-independent power-law exponent. We do not expect qualitative changes of this scenario by further refinements of the model, like partly fixing scaling factors for the modeling of level-scaling or by changing the number of branches \cite{KeaDwyBre97}.

In order to draw conclusions from our model about Hamiltonian systems, we make use of the hypothesis that there exists a universal mechanism generating the hierarchical fine-scale structure of Hamiltonian systems, leading to a universal distribution of scaling factors. In particular, we consider different Hamiltonian systems, with respect to their fine-scale structure, as different realizations from one statistical ensemble, just as this is the case for a specific realization of the random scaling factors $A$ and $B$ in the tree model. The immediate consequence of this hypothesis and of the realization independence of the power-law decays found in our model is, that all Hamiltonian systems should exhibit the same universal power-law decay. The numerically observed different power-law decays over finite times are merely realization-dependent fluctuations around the universal decay. 
 
As another important consequence, we are now able to effectively address the practical question of how to numerically determine the universal power-law exponent.  The straightforward approach is to extract it from $P(t)$ of a single Hamiltonian system determined up to very large times.  However, in order to have a reliable power-law exponent from a single  $P(t)$ one would need much larger times than in previous studies (which would go beyond current computational resources). The reason is that the trajectories of the chosen Hamiltonian system must have explored a statistically significant sample of the fine structure of its phase-space in order to sample the universal distribution of scaling factors. Instead, we propose to average $P(t)$ over different Hamiltonian systems. This procedure averages  out fluctuations and should give a much better defined power-law decay already for smaller times, see e.g.\ Refs.~\cite{ChiShe81,ChrGra93}. It is important to stress, that due to the results of our model such an averaging has a clear meaning: It is not an average over different power laws, but it corresponds to a faster reduction of the realization-dependent fluctuations around the universal decay, obtained by an increased statistical sampling of the universal distribution of scaling factors.

\begin{figure}[t]
{\epsfxsize=\columnwidth \epsfbox{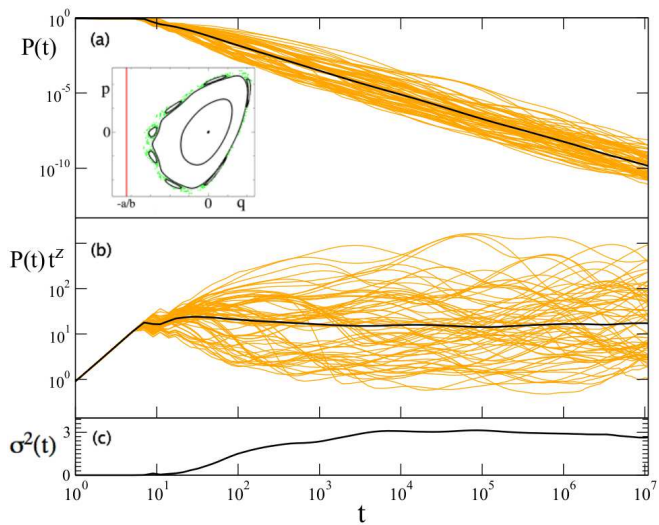}\label{real}}
\vglue -0.2cm
\caption{(Color online) (a) Survival probabilities $P(t)$ for $100$ area preserving maps (thin lines), defined in the text, and their logarithmic average (thick line). Inset: phase space of one of these maps, showing the main island and the border line to the escape region.
(b) Rescaled survival probabilities  $P(t) \, t^{z}$ with the fitted exponent $z=1.57$ from (a).
(c) Variance $\sigma^2(t)$ of $\ln [P(t) \, t^{z}]$ showing saturation.
}
\label{fig3}
\end{figure}

Guided by this idea we study an ensemble of area preserving maps of the form $p_{n+1}=p_n -V'(q_n)$; $q_{n+1}=q_n+p_{n+1}$ with $V(q)=\frac{a}2q^2+\frac{b}3q^3$ for 100 pairs of parameters $(1<a,b<2)$. These maps have a stable fixed point at $(0,0)$ and an unstable fixed point at $(-\frac{a}b,0)$. The fine scale structure at the border of the regular island around the stable fixed point is quite different for all realizations. We start $10^{11}$ trajectories uniformly in a box next to the unstable fixed point, $q\in[-\frac{a}b,-\frac{a}b+ \delta], p\in[-\delta,\delta]$ with $\delta=0.01$, and choose the escape region  $q\le-a/b$. Fig.~\ref{fig3} shows the individual survival probabilities $P(t)$ together with their average.
Note, that the invariant initial condition in the tree model corresponds in the Hamiltonian system to a uniform initial distribution just in the chaotic component. Numerically, this initial condition is not achievable. Instead, we choose an initial condition in a box in the chaotic region far away from the island, which changes the exponent by one, $z=\ztree+1$~\cite{Mei96}.

The averaged curve in Fig.~\ref{fig3} presents a well defined power-law decay with exponent $z \simeq 1.57\pm 0.03$, where the error describes the variation for different fitting ranges. Quite importantly, the variance of the fluctuations around this averaged power-law saturates. This agrees with the prediction of our model (see Fig.~\ref{fig1}c) and it gives direct evidence that all individual power-law decays are consistent with a single universal decay.
We  find similar results with the same exponent for an ensemble of kicked double-harmonic potentials. Motivated by our work, Altmann \cite{Altmann} studied an ensemble of modified standard maps, confirming our results qualitatively and quantitatively.

In conclusion, we conjecture a universal power-law decay $P(t) \sim t^{-z}$ from the analysis of a Markov model with random scaling factors and from the hypothesis of universality in the statistical properties of the fine-scale structure of typical Hamiltonian systems.  Our numerical investigations support this hypothesis and suggest the universal exponent $z \simeq 1.57\pm 0.03$. Future studies will concentrate  on a direct extraction  of the  universal distributions of area and escape-rate scalings in  generic  Hamiltonian systems (both numerically and analytically),  that  will eventually  lead to a fully theoretical derivation of  the universal exponent $z$.

We thank E.G.Altmann for useful discussions.
This research was supported in part by
the DFG under contract KE537/3-3 and
the NSF under Grant No. PHY05-51164.

\end{document}